\documentclass[twocolumn,preprintnumbers,amsmath,amssymb, showpacs]{revtex4}
\usepackage[utf8]{inputenc}
\usepackage{graphics}
\usepackage{graphicx}% Include figure files
\usepackage{psfrag}
\usepackage{amsmath}
\usepackage{upgreek}
\usepackage{empheq}
\usepackage{verbatim}

\DeclareSymbolFont{UPM}{U}{eur}{m}{n}
\DeclareMathSymbol{\uppartial}{0}{UPM}{"40}
\DeclareMathAlphabet\matheuler{U}{eur}{m}{n}
\newcommand{\ee}{\ensuremath{\matheuler{e}}} % Euler Roman "e" als Euler-Zahl
\newcommand{\ii}{\ensuremath{\matheuler{i}}} % Euler Roman "i" als imaginäre Einheit.
\usepackage{isotope}
\usepackage{hyperref}
     \usepackage{mathptmx}
     \usepackage[scaled=0.92]{helvet}
     \usepackage{courier}
\begin{document}

%\preprint{TUD-ITP-TQO/04-2015-V100303}

%\title{ Cooling and frequency change of an impurity in a superfluid\\using an open system approach}
%\title{Damping rate, frequency shift, and memory effects of a quantum oscillator in an ultracold Bose gas}
\title{Cooling and frequency shift of an impurity in a ultracold Bose gas\\using an open system approach}

\author{Paula Ostmann}
     \affiliation{Institut f\"{u}r Theoretische Physik,
Technische Universit\"at Dresden, D-01062 Dresden, Germany}
\author{Walter T. Strunz}
   \affiliation{Institut f\"{u}r Theoretische Physik,
Technische Universit\"at Dresden, D-01062 Dresden, Germany}

\date{\today}

\begin{abstract}
We investigate the quantum dynamics of a harmonically trapped particle (e.g. an ion) that is immersed in a Bose--Einstein condensate. 
The ultracold environment acts as a refrigerator, and thus, the influence on the motion of the ion is dissipative. We study the 
fully coupled quantum dynamics of particle and Bose gas in a linearized regime, treating the quasi-particle excitations of the gas
as a (non-Markovian) environment for the particle dynamics. The density operator of the latter follows a known non-Markovian master
equation with a highly non-trivial bath correlation function that we determine and study in detail. The corresponding
damping rate and frequency shift of the particle oscillations can be read off. We are able to identify a Quantum Landau 
criterion for harmonically trapped particles in a superfluid environment: for frequencies 
$\omega$ well below the chemical potential, the damping rate is strongly suppressed by a power law $\omega^{{4}}$.
This criterion can be seen as emerging from the classical Landau criterion involving a critical velocity combined with
Heisenberg's uncertainty principle for the localized wave packet of the quantum particle.
Furthermore, due to the finite size of the Bose gas, after some time we observe memory effects and thus non-Markovian 
dynamics of the quantum oscillator.

\end{abstract}
\pacs{03.65.Yz, 05.40.Jc, 37.10.Rs, 67.85.De}
\maketitle
%\section{Introduction}
$Introduction$-
Experiments immersing single ions or particles into Bose--Einstein condensates (BEC) is a very active and ambitious field in the physics of cold gases. 
The first experiments with ions \cite{zipkes_trapped_2010, schmid_dynamics_2010, hall_light-assisted_2011, hall_millikelvin_2012, ravi_cooling_2012, 
sivarajah_evidence_2012, deiglmayr_reactive_2012}, or neutral atoms \cite{spethmann_dynamics_2012} already offered a great opportunity to 
study properties of ultracold gases as well as of ions and their interaction. Moreover, new techniques and experimental setups provide a 
wider scope in studying hybrid systems \cite{balewski_coupling_2013,huber_far-off-resonance_2014}.
Furthermore, such experiments offer a playground for studying the quantum dynamics of a single particle coupled to a superfluid environment.

Theoretically, the ground state properties of the composed system already have been studied \cite{dalfovo_f_condensate_1996, cote_mesoscopic_2002, massignan_static_2005, 
schurer_ground-state_2014, akram_numerical_2016}. For the dynamical interaction between the two components different 
approaches have been used, ranging from collisions between ion and condensate particles \cite{cote_ultracold_2000,idziaszek_controlled_2007,rellergert_measurement_2011,zipkes_kinetics_2011,
tomza_cold_2015} over three-body processes to master equation approaches \cite{klein_dynamics_2007,krych_description_2015}, or the MCTDHB method \cite{schurer_capture_2015, kronke_correlated_2015}. Considering the BEC as an environment has also 
been proven useful in different contexts such as lattice setups \cite{klein_interaction_2005, johnson_impurity_2011} or in polaron-type 
approaches \cite{casteels_many-polaron_2011, ardila_impurity_2015,lampo_bose_2017}.

The standard approach to the quantum dynamics of a single particle and its many-degree of freedom environment is to derive a master equation 
for the open system. The properties of the bath and its interaction with the open system lead to damping terms and frequency shifts (e.g. the
lamb shift) often involving a Golden-rule damping rate in the weak coupling and Markovian limit. This damping rate (and the frequency shift)
can be determined from the underlying bath correlation function. Thus, once the latter is determined, the quantum dynamics of the single
particle can be determined from a master equation.

%or a fixed spectral density distribution. Instead of using this access we calculate the spectral density of BEC. With this information it is already possible to gain a detailed impression of the dynamics of the ion in the BEC, without solving the actual dynamics.\\
We start from the underlying full Hamiltonian of particle and Bose gas and assume the latter to be so cold ($T=0$) that a
condensate wave function $\Psi_0(\vec r)$ can be identified. In the usual way we then linearize and study the fully quantum particle+BEC 
dynamics. Crucially, the zero temperature bath correlation function $\alpha\left(t-s\right)= \int_0^\infty d\omega J(\omega)\ee^{-\ii\omega\left( t-s\right) }$ with its spectral 
density $J\left(\omega\right)$ has to be determined (see also \cite{klein_interaction_2005}). In spectral representation, this amounts to
determining the quasi-particle Bogoliubov spectrum and corresponding wave functions from the corresponding Bogoliubov-de Gennes equation, 
a formidable task. Here, however, we present a simple direct 
route to determine $\alpha(t-s)$ in the time domain using the Gross--Pitaevskii equation (GPE) \cite{gross_hydrodynamics_1963}. We stress that 
this does {\it not} imply any mean-field approximation but exploits the equivalence of quantum and classical dynamics for linear systems. 
The numerically determined bath correlation function thus obtained shows that the dynamics can be divided in an initial dissipative oscillation 
at a shifted frequency. At a later time, however due to the finite size of the trapped BEC a small fraction of the energy may return to the 
oscillator, which was also observed in the T-matrix approximation \cite{Volosniev_real-time_2015}. Hence the effect of the environment on the 
particle dynamics is three-fold: frequency shift, damped dynamics and a possible reheating, that shall be discussed in the last section.
\newline
%Damping is a omnipresent process which is mostly just modeled as a fitted damping rate or back round noise to simulate the energy loss of the ion into the bosonic bath. Still this process is not understood in detail (Quelle?). In this article we want to gain deeper insight into the understanding of the damping process. By combining two approaches to the problem we do not just find detailed information about the dynamics of a  quantum particle in contact to a bath, but also we can study properties of such a bath by using considering the effect of the ion on the BEC-environment. \\
%The standard approach to the problem of a single quantum system and a macroscopic environment is to derive the master equation for the system by including the bath as a damping term described by a damping rate. This rate is usually interpolated with the help of experimental data (Quelle?). Instead of suing this approach we calculate the damping rate by using the spectral density of the BEC. With this density it is already possible to gain a detailed impression of the dynamics of the ion in the BEC an furthermore we can deduce about other phenomena. On the one hand we can reason about superfluidity in BEC and question the validity of the Landau criterion. On the other hand the finiteness of the bath itself offers an opportunity for a comeback of the initial state which would result in a periodically cooling an reheating of the ion over the course of time.\\
%\section{Full Model and Linearisation}\label{model}

$Full~Model~and~Linearisation$- For describing a single quantum particle in a BEC we start with the following Hamilton operator:
\small{
\begin{align}\label{Hamilton}
& \hat H  = \frac{\hat{\vec{P}}^{\,2}_{\vec{R}}}{2 M } +  \frac{1}{2}M\Omega_0^2{\hat{\vec{R}}}^2 +\int\limits_{-\infty}^{\infty} \mathrm{d}^3 r~~V
\left(\vec{r},\hat{\vec{R}}\right)\hat\Psi^\dagger\left(\vec{r}\right) \hat\Psi\left(\vec{r}\right) \\ \nonumber
&\int\limits_{-\infty}^{\infty} \mathrm{d}^3 r \Big( -\hat\Psi^{\dagger}\left(\vec{r}\right)\frac{\hbar^2}{2 m}\nabla^2\hat \Psi\left(\vec{r}\right)
+ U\left(\vec{r}\right)\hat\Psi^\dagger\left(\vec{r}\right) \hat\Psi\left(\vec{r}\right)\\\nonumber
&+ \frac{g}{2}\hat\Psi^\dagger\left(\vec{r}\right) \hat\Psi^\dagger\left(\vec{r}\right) \hat\Psi\left(\vec{r}\right)\hat\Psi\left(\vec{r}\right) \Big) \nonumber
\end{align}}
\normalsize
Here, the ion (the quantum particle) is described by a position $\hat{\vec R}$ and corresponding momentum $\hat{\vec P}$ operator.
It is trapped in a harmonic potential with bare trapping frequency $\Omega_0$. The Bose gas is described by the bosonic field operators 
$\hat \Psi\left(\vec{r}\right)$ and $\hat\Psi^\dagger\left(\vec{r}\right)$. The first integral is the interaction energy between 
ion and Bose gas. For actual calculations shown later, we use a cut-off polarization potential 
$V\left(\vec{r},\vec{R}\right) = \frac{V_0}{\left(1 + \left|\vec{r}- \vec{R}\right|^2a^{-2}\right)^2}$, introducing the interaction 
energy scale $V_0=\frac{\tilde{V}_0}{a^4}$ and length scale $a$. The second integral is the energy operator of the Bose gas with a 
harmonic
trap potential $U\left(\vec{r}\right)=\frac{1}{2}m\omega_{\rm BEC}^2{\vec r}^2$, 
$g=\frac{4\pi\hbar^2 a_{\mathrm s}}{m}$ is the interaction strength between the Bose particles
with $a_{\mathrm s}$ the s-wave scattering length.
Obviously, for matters of simplicity, we assume the two traps (of the particle and of the gas) have their minimum at the same position 
${\vec R}_0={\vec r}_0=\vec 0$.

For all that follows we assume that the ion has already been cooled down close to its ground state and its motion is
restricted to small oscillations around its equilibrium position. Thus, energy loss from the ion does not 
result in an atom loss out of the BEC. This assumption is being further incorporated by performing a Taylor expansion around the 
equilibrium position $\vec{R}_0 = 0$. Moreover, we assume a well-occupied condensate 
wave function ($T=0$) such that the usual symmetry breaking ansatz for the field operator $\hat\Psi = \Psi_0 +\hat{\delta \Psi}$ is 
well justified.
As usual, we will truncate the expansion of the full Hamiltonian after the quadratic terms, leading to a quasi-particle spectrum for
the excitations of the ultracold gas, influenced by a (static) contribution of the ion-gas interaction. There is also a static
contribution to the particle potential which is just the interaction potential averaged over the condensate
wave function. In our Taylor expansion, this is the first (static) contribution to a frequency shift of the ion dynamics,
\begin{equation}\label{static_shift}
 \Omega_0^2 \rightarrow \Omega^2 = \Omega_0^2 + \delta\Omega_0^2.
\end{equation}
This frequency shift can be determined analytically employing the Thomas--Fermi approximation for the condensate wave function, giving $\delta\Omega_0^2 = \frac{\pi}{32}\frac{a}{a_\mathrm{s}}
V_0^2 \frac{m}{M}$ and neglecting terms that are smaller by a factor $(1/N)$.
The change in frequency is confining, meaning the ion is trapped more tightly. The correction 
is so strong that we expect the ion to stay in the condensate even after turning off the external ion trap. Remarkably, the correction is also confining for a repulsive interaction due
to the quadratic dependence on $V_0$. 

Dynamically, the interaction of the ion with the gas will then lead to an excitation of these quasi-particles and thus
to a damping of its oscillatory motion: the ion will be further cooled.
%$\delta\Psi\left(t\right)$ that is sufficiently small in comparison to the static part of the bath $\Psi_0$.\\
By performing the linearization, we arrive at a non-diagonal quadratic Hamilton operator. Purely linear terms in $\hat{\delta \Psi}$ 
vanish by minimizing the Hamiltonian with respect to the condensate wave function $\Psi_0$ whose resulting GPE contains a static contribution from the ion-gas interaction. 
%The $\delta\Psi\left(t\right)$ - independent part can be written as an integral over the Gross-Pitaevskii energy which we affiliate to $\hat H_{\mathrm{ion}}$ since it is a constant for the rest of the discussion. 
Using a Bogoliubov transformation \cite{bogolubov_theory_1946} we arrive at a standard Hamilton operator for a quantum
harmonic oscillator, coupled to a bosonic bath. This is sometimes referred to as a quantum Brownian motion model \cite{haake_strong_1985, hu_quantum_1993, strunz_convolutionless_2004}:
% \begin{comment}$ \delta \hat{\Psi}\left(\vec{r}\right)= \sum\limits_{\nu=0}^{\infty} \left[ u_\nu\left( \vec{r} \right) \hat{b}_\nu \ee^{-\ii \omega_\nu t} +  v^*_\nu\left( \vec{r} \right) \hat{b}^\dagger_\nu \ee^{\ii \omega_\nu t}\right]$ \end{comment} 
\small
\begin{align}\label{H_final}
%\hat {H}  &= \frac{\hat{\vec{P}}^{\,2}_{\vec{R}}}{2 M } +  \frac{M}{2}\Omega^2\left(\vec{R}-\vec{R}_0\right)^2 +  \sum_\lambda \hbar \omega_\lambda \hat{b}^\dagger_\lambda \hat{b}_\lambda\nonumber\\
\hat {H}  = \frac{\hat{\vec{P}}^{\,2}_{\vec{R}}}{2 M } +  \frac{M}{2}\Omega^2\hat{\vec{R}}^2 +  \sum_\lambda \hbar \omega_\lambda \hat{b}^\dagger_\lambda \hat{b}_\lambda 
+ \hat{\vec{R}}\cdot \sum_\lambda \left( \vec{g}_\lambda \hat{b}^\dagger_\lambda +  \vec{g}^*_\lambda \hat{b}_\lambda    \right) 
\end{align}
\normalsize
Here, $\Omega$ is the shifted frequency, $\{\hbar\omega_\lambda\}$ is the Bogoliubov spectrum and $\vec{g}_\lambda$ are coupling vectors determined from the
interaction potential $V(\vec r,\vec R)$, the condensate wave function $\Psi_0(\vec r)$, and the solutions of the
Bogoliubov--de Gennes equation.
 \begin{comment}$ \vec{g}_\lambda= \int\limits_{-\infty}^{\infty} \mathrm{d}^3 r \left( \Psi_0\left( \vec{r} \right)v_\lambda\left( \vec{r} \right) + \Psi^*_0\left( \vec{r} \right) u_\lambda\left( \vec{r} \right)\right)\nabla_{\vec{r}'} V \left(\vec{r},\vec{r}'\right)|_{\vec{r}_0}$\end{comment}
%\cite{astrakharchik_motion_2004, gribakin_calculation_1993, molof_measurements_1974}\\
%
%\section{The bath correlation function}
\newline

$Bath~correlation~function$- 
The linear model (\ref{H_final}) has been studied in detail. The influence of the environment onto the central oscillator
can be fully captured by the bath correlation function $\alpha\left( t-s\right)$, which is a force-force correlation function
entirely determined by the $\omega_\lambda$ and $\vec{g}_\lambda$. For our three dimensional oscillator, the bath correlation function is
a tensor, reading (at zero temperature considered here), 
\begin{equation}\label{Alpha_}
 \underline{\underline{\alpha}}\left(t-s\right) =  \sum\limits_{\lambda=0}^{\infty} \left(\vec{g_\lambda} \otimes \vec{g_\lambda}^*\right) \ee^{-\ii \omega_\lambda \left( t-s \right)}
\end{equation}
%We will furthermore consider just the first matrix element of $ \underline{\underline{\alpha}}\left(t-s\right)$, due to the spherical symmetry which will be applied to the system.
According to the derivation of ($\ref{H_final}$) $\omega_\lambda$ and $\vec{g_\lambda}$ can be calculated from the Bogoliubov--de Gennes equation. 
Alternatively, and this turned out to be a much more direct route,
%However this straight-forward method will not result in an exact solution due to the truncation in $\lambda$, which needs to be performed.\\
we are able to construct the imaginary part of $\alpha\left(t-s\right)$ from an autocorrelation function (set $s=0$) of (a vector of) 
wave functions $\delta\Psi(\vec r,t)$:
\small
\begin{equation}\label{beta}
 2\mathfrak{Im}{\underline{\underline\alpha}}\left(t\right) = \frac{1}{\left|\epsilon\right|^2}\left(\langle{\vec{\delta 
\Psi}}\left(0 \right)|\otimes|{\vec{\delta\Psi}}\left(t \right)\rangle - \langle{{\vec{\delta \Psi}}\left(t \right)|
\otimes|{\vec{\delta \Psi}}\left(0 \right)}\right)
\end{equation}
\normalsize
Here, the initial condition for the fluctuation ${\vec{\delta\Psi}}_0\left( 0\right) $ needs to be chosen as
\begin{equation}\label{initial}
 \vec{\delta \Psi}_0\left(\vec{r}\right) = -\ii \epsilon \Psi_0\left(\vec{r}\right)\vec\nabla_{\vec{R}} V \left(\vec{r},\vec{R}\right)|_{\vec{R}_0}
\end{equation}
with $\epsilon$ a small parameter to ensure the dynamics to stay in the linearized regime.
The time evolution of $\delta\Psi_0\left( t\right) $ can be obtained using the GPE for $\Psi\left( t\right) $ and subtracting the ground state $\Psi_0$. It is important to note, that this does not correspond to a mean field approximation, it merely is a practical
mean to circumvent an explicit solution of the Bogoliubov-de Gennes equation. Additionally it must not be forgotten, that the time evolution of $\delta\Psi\left( t\right) $, does not correspond to the real dynamics of the ion, it just shows the propagation of a perturbation, leading to the correlation dynamics $2\mathfrak{Im}{\underline{\underline\alpha}}\left(t\right)$. 

%A more intuitive function to look at is the spectral density, which is the one-sided Fourier transform of $\alpha\left( t-s\right)  =  \int\limits_{0}^{\infty} \mathrm{d} \omega  J\left(\omega\right) \ee^{-\ii \omega \left( t-s\right) }$. In the weak coupling limit $J\left(\omega\right)$ can be connected via Fermi's golden rule \cite{dirac_quantum_1927} to a classical damping rate $\gamma = \pi \frac{J\left(\omega\right)}{\omega}$. Yet in the strong coupling limit it still displays the damping strength. By symmetry considerations the full bath correlation function can be constructed.\\ 
%\section{Numerical Implementation for a single Ion in a BEC}

We consider a three dimensional system consisting of a BEC and a single ion. As parameters for our calculation we use a spherically trapped BEC (frequency $\omega_{\mathrm{BEC}}$, that will
be used as the fundamental scale for all quantities). We choose an
interaction energy scale $V_0=\hbar\omega_{\mathrm{BEC}}$. The positively charged ion is 
placed in the center.
%\begin{figure}[h!]
% \begin{center}	
%    \includegraphics[width=.4\textwidth]{bec.eps}
     
\begin{figure}[h!]
  \begin{center}	
    \includegraphics[width=.45\textwidth]{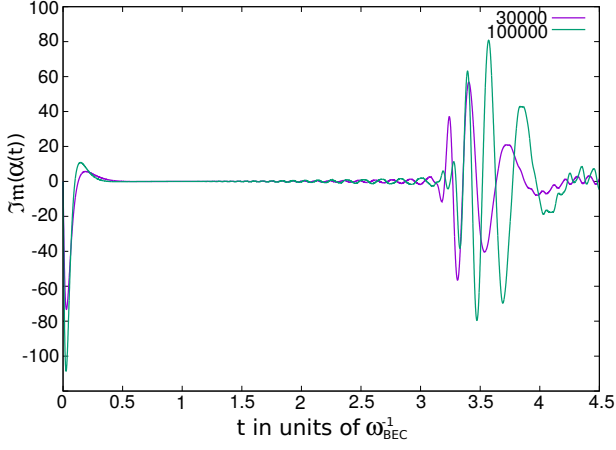} 
  \caption{The imaginary part of $\alpha\left( t\right) $ was calculates for two different particle numbers in the BEC using ($\ref{beta}$). The initial energy decay as well as the energy return can be seen in the correlation dynamics.}\label{beta_pic}
 \end{center}
\end{figure}

%  \caption{Shown is a cut through the time evolved perturbation $\left|\delta\Psi_0\left( t\right) \right|^2$ in the $z=0$-plane at time $t=0.5$ in a BEC. The process of energy loss can be seen as sound waves propagating through the condensate. } \label{ini_state}
% \end{center}
%\end{figure}
Due to the spherical symmetry, the tensor bath correlation function becomes a scalar
$\alpha(t-s)$ times the $(3x3)$-identity matrix.
We calculate the ground state $\Psi_0$ for the system using imaginary time propagation \cite{kosloff_direct_1986}. Afterwards $\Psi\left( t\right) = \Psi_0 + \delta\Psi\left(t\right) $ is computed by solving the GPE using the Split-Operator method and the initial condition ($\ref{initial}$), with $\epsilon=0.005$. It can be seen in the simulations that the initial perturbation ($\ref{initial}$) decays into the BEC via spherical sound waves. Propagating further in time we observe a return of energy due to reflection at the border of the BEC, as illustrated in Figure $\ref{beta_pic}$ showing $2\mathfrak{Im}\alpha\left(t\right)$ for two different particle numbers.
%\vspace{-1cm}
It is interesting to note that the return time $T_{\mathrm{ret}}$ is independent of the particle number, as well as particle species or other parameters, only the trapping frequency of the BEC enters, as it was already observed in \cite{Volosniev_real-time_2015}. A simple estimate of the return time
is based on a sound wave traveling through a spherically symmetric BEC in Thomas--Fermi approximation:
\begin{align}\label{speedofsound}
 T_{\mathrm{ret}} \approx 2 \int\limits_0^{R_\mathrm{TF}} \frac{d r}{c_{\mathrm{sound}}\left(r\right)}\;\;
 \mbox{with}\;\;
c_{\mathrm{sound}}(r) = \sqrt{\frac{g\left|\Psi\left(\vec{r}\right)\right|^2}{m}}
\end{align}

The integral is easily calculated giving $T_{\mathrm{ret}} \approx \sqrt{2}\frac{\uppi}{w_{\mathrm{BEC}}}$, indeed, independent of the particle number. This means that the
time of returning energy can be controlled by opening or closing the BEC trap. The effect, to which we refer to as coherent heating, shall be discussed in the last part of the article.\\
\newline
%\section{Open system dynamic}
$Open~system~dynamics$- 
The dynamics of the ion can be described using the well known master equation for a quantum 
brownian particle in a harmonic oscillator potential with 
$\hat H_{\mathrm S}$ from equ. (\ref{H_final}), which can be derived using different approaches \cite{haake_strong_1985,hu_quantum_1993,strunz_convolutionless_2004}, it can be written as 
\small
\begin{align}\label{master_ion_eq}
 \partial_t\rho_t &= -\ii\left[\hat H_{\mathrm{S}}, \rho_t  \right] - \ii A\left( t\right) \left[\vec{R}^2, \rho_t\right] - \ii B\left( t\right)\left[\vec{R}, \left\{\vec{p}, \rho_t\right\}\right]\nonumber\\
&+C\left( t\right) \left[\vec{R},\left[\vec{p}, \rho_z\right]\right]- D\left( t\right) \left[\vec{R}, \left[\vec{R}, \rho_t\right]\right].
%\\
%A\left( t\right) &= \mathfrak{Im}\left( \int\limits_0^t \mathrm{d}s~\alpha\left( t-s\right) \left( F\left( t,s\right) +H\left( t,s\right) \right) \right) \nonumber
\end{align}
\normalsize
Here, $A\left( t\right) , B\left( t\right) , C\left( t\right) , D\left( t\right)$ are 
known complicated functions depending on the bath correlation function $\alpha\left( t-s\right)$.
They assume constant values once the bath correlation function has decayed to zero. 
The term involving $A$ amounts to an additional (Lamb) frequency shift that we can neglect
compared to the earlier static shift from (\ref{static_shift}) in the parameter regime 
presented here. The other terms describe damping ($B$) and quantum diffusion ($C$ and $D$).

Instead of solving for the full density operator of the ion, in the following we will simply
determine its position expectation value $\vec{Q}=<\vec{R}>={\mathrm{tr}}\left(\vec{R}\hat\rho_t\right)$, whose equation of motion
we can also obtain directly from equ. (\ref{H_final}).

%\small
%begin{equation}\label{preS-tep_cl_eq}
% M\ddot {\vec{Q}} \left( t\right)  -M \Omega^2 \vec{Q}\left( t\right)  - \frac{2}{\hbar}  \int \limits_0^t \mathrm{d}s \mathfrak{Im}\left(\underline{\underline{\alpha}}\left( t-s\right)  \right) \vec{Q}\left( s\right) =0
%\end{equation}
%\normalsize
By introducing the damping kernel $\Gamma(t-s)$ through 
$\frac{2}{M\hbar} \mathfrak{Im}\alpha\left( t-s\right) = 
- \frac{\uppartial}{\uppartial s} \Gamma \left( t-s\right)$ and partial integration, we get
the evolution equation for a damped harmonic oscillator with memory
\begin{equation}\label{cl_delay_1}
 \ddot {\vec{Q}}  + \Omega^2 \vec{Q}  + \int\limits_{0}^{t} \mathrm{d}s \Gamma\left(t-s\right) \dot {\vec{Q}}  \left(s\right) = 0,
% {\vec{Q}}  \left(s\right) = \Gamma\left(t\right)\vec{Q}\left(0\right).
\end{equation}
where we assume initial conditions $\vec Q(0)=0$, $\dot{\vec Q}(0)=(1,0,0)$.
The solution for $\vec{Q}\left( t\right) $ can be computed using a Laplace transform of equation ($\ref{cl_delay_1}$) (see \cite{weiss_quantum_2008}).

\begin{figure}[h!]
  \begin{center}	
    \includegraphics[width=.45\textwidth]{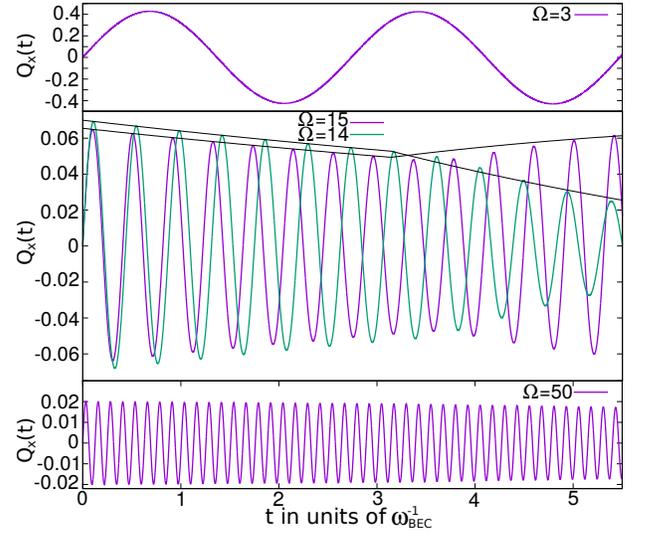} 
  \caption{The solution $\vec{Q}\left( t\right)$ is shown for three different frequencies. The initial conditions where chosen with a finite velocity only. Additionally the exponential decay rate is fitted, which refers to solution ($\ref{delay_eq}$).}\label{Klassik_50}
 \end{center}
\end{figure}

The ion dynamics for three different frequencies ($\Omega/\omega_{\mathrm{BEC}}=3,15,50$) are illustrated in Figure $\ref{Klassik_50}$ where show the non-vanishing $Q_x$-component.
%As can be seen the damping, as well as the recurrence strength depend strongly on the frequency. \\
We see a very small damping for low and high frequencies ($3,50$) and a more significant decay
for an intermediate frequency $\Omega/\omega_{\mathrm{BEC}}=15$. Remarkably, the return of the
bath correlation for times $\omega_{\mathrm{BEC}}t\approx\pi$, as visible in Fig. (\ref{beta_pic}),
has a major impact on the ion dynamics: memory effects (non-Markovian dynamics) due to the finite
size of the BEC become clearly important. With a surprisingly sensitive dependence on the
oscillator frequency $\Omega$, we can see heating (for $\Omega = 15\omega_{\mathrm{BEC}}$)
or additional cooling (for $\Omega = 14\omega_{\mathrm{BEC}}$). This phenomenon will be explained
in more detail later.

For now, let us concentrate on determining the initial damping rate. That initial dynamics
would also be seen in an infinite environment, when the returns of the bath correlation function
are neglected (i.e. we set $\alpha(t)=0$ for $\omega_{\mathrm{BEC}}t\ge 1$ in 
Fig. (\ref{beta_pic})).

\begin{figure}[h!]
  \begin{center}	
    \includegraphics[width=.44\textwidth]{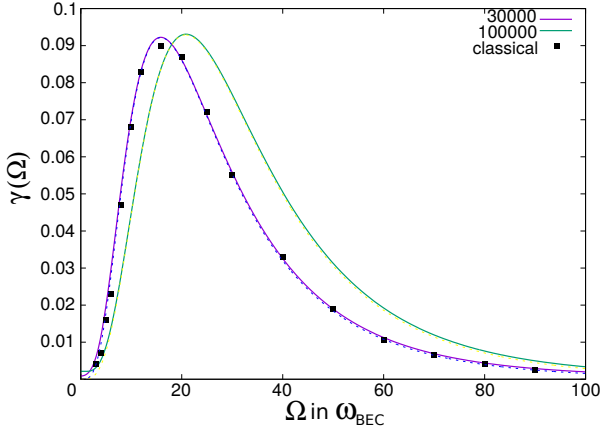} 
  \caption{The plot shows the damping rate $\gamma\left(\Omega\right)$ for two different particle numbers. The solid lines represent the numerically obtained result based on the spectral density.
  The black squares are rates fitted to the solutions of the classical equation of motion $(\ref{cl_delay_1})$ as shown in Fig. (\ref{Klassik_50}). The dashed lines, indistinguishable from the numerical result, show an analytical expression based on a homogeneous gas, as described by ($\ref{J_explicit}$).}\label{J_ini}
 \end{center}
\end{figure}

The cooling (for small damping) is most easily determined from
the Laplace transform of $\Gamma(t)$ (neglecting the returns) whose half
real part determines the damping rate $\gamma(\Omega) = \frac{\pi}{M\hbar+}\frac{J(\Omega)}{\Omega}$ of the oscillator with frequency $\Omega$ (Fermi's golden rule).
This $\gamma(\Omega)$ is displayed in Fig. (\ref{J_ini}). In full agreement with
the numerical solution of $Q(t)$ from equ. (\ref{cl_delay_1}) (black dots),
$\gamma(\Omega)$ tends to zero for small and large frequencies and has a clear maximum for
intermediate values.\\
%the influence of t
%From equation ($\ref{cl_delay_1}$) we can also see, that there are just two determining parameters for the ion dynamic: the frequency $\omega$ and the from $\alpha\left( t-s\right) $ deduced $\Gamma\left( t-s\right) $. Similar to $J\left(\omega\right)$ it is more convenient to look at $\tilde\Gamma\left(\omega\right)$, shown in picture $\ref{J_ini}$.
$Quantum~Landau~criterion$- 
To understand the damping rate $\gamma(\Omega)$ of the ion more thoroughly, we discuss the structure of the spectral density $J\left(\omega\right)$ by calculating it for a homogeneous BEC, representing an infinite environment for the ion. In this case the solutions of the Bogoliubov--de Gennes equations are known \cite{pitaevskii_bose-einstein_2003}. 
By using the corresponding plain wave expansion for $\delta \Psi\left( t\right) $ it is 
possible to compute a closed expression for $\gamma(\Omega)$:
%\small
%\begin{equation}
% \alpha\left(t\right) = \frac{\mu}{Ng} \left(\frac{1}{2\pi} \right)^3 \frac{4\pi}{3} \int \mathrm{d} k k^4 \left| \tilde V\left(\vec{k}\right)\right|^2 \left(u_k-v_k\right)^2 \ee^{-\ii\omega t}
%\end{equation}
%\normalsize
%for the specific polarization potential $J\left(\omega\right)$ can be derived:
%\begin{align}\label{J_explicit}
% J\left(\omega\right) = \left(2\pi\right)^3 \frac{4\pi}{3} \pi^4 \sqrt{2}^3 \frac{\bar{V}_0^2\bar{n}_0}{\bar{a}^2}\frac{\hbar}{\sqrt{m}}\frac{1}{\omega_{\mathrm{BEC}}}\\  \frac{\sqrt{\sqrt{\omega^2+\mu^2}-\mu}^5}{\sqrt{\omega^2+\mu^2}} \ee^{-2a\sqrt{2\sqrt{\omega^2+\mu^2}-\mu}} \nonumber
%\end{align}
\begin{align}\label{J_explicit}
 \gamma\left(\Omega\right) &= C \frac{k(\Omega)^5}{\Omega\sqrt{\hbar^2\Omega^2+\mu^2}}
 |\tilde{V}\left(k(\Omega)\right)|^2 ~~~\mathrm{with}\\
 k(\Omega)&=\sqrt{\frac{2m}{\hbar^2}\left(\sqrt{\hbar^2\Omega^2+\mu^2}-\mu\right)}.\nonumber
\end{align}
$C$ summarizes all prefactors as $C= \mu(12MNg\pi^2)^{-1}$. Moreover, ${\tilde V}(k)$ is the Fourier transform of the (isotropic) interaction
potential $V(r)$. 
This approximate $\gamma(\Omega)$ from (\ref{J_explicit}) 
is shown in Figure $\ref{J_ini}$ as dashed lines. For this set of 
parameters the solutions coincide exceptionally well with the numerically obtained graph.

Result (\ref{J_explicit}) explains how the decay of the
damping rate for high frequencies is directly determined from the large $k$-dependence
of ${\tilde V}(k)$ and is thus a specific property of the ion-gas interaction.
By contrast, the low-frequency behavior $\Omega\rightarrow 0$ is independent
of the details of the interaction. For $\hbar\Omega\ll \mu$, we have $k(\Omega) \sim \Omega$
and therefore $\gamma(\Omega)\sim \Omega^4$. Therefore, damping in the low 
frequency limit is strongly (and universally) suppressed.

%Note that the first part in $\omega$ originates from $u_k$ and $v_k$, while the second part is due to the Fourier transform of the interaction potential $\tilde V\left(k\right)$. Hence the decrease of the damping rate for high ion frequencies is a result of the interaction potential and is thus depending on the impurity (i.e. ion, atom, molecule etc.). The solution is shown in picture $\ref{J_ini}$  as the dashed lines.For this set of parameters the solutions coincide exceptionally well with the numerically obtained graph. Nevertheless for stronger interactions the simple homogeneous picture deviates more from the physical correct assumptions, making the numerical evaluation more deciding.\\
%From the graphs for the different particle numbers it is clear to see that with higher particle number in the BEC also a higher damping rate for the ion occurs. For all particle numbers we find, that there is am initial growth of the the damping rate for small frequencies until a maximum is reached. Afterwards $J\left(\omega\right)$ decays exponentially. This infers that there is a trapping frequency regime for the ion in which the cooling process is by orders of magnitude more efficient than in the higher frequency regime, which would be the usually experimentally excessed paul trap frequencies (VERWEIS).\\
%As can be seen in $\tilde \Gamma\left(\omega\right)$ in the low frequency regime, damping is strongly suppressed. he classical trajectory for $\omega=3$ shows a lack of dissipation as well. 
This behavior shall be interpreted in terms of superfluidity of the BEC for the motion of a 
quantum particle,  introducing a Quantum Landau Criterion.\\ In the classical Landau Criterion a critical velocity $v_{\mathrm{crit}}$ is introduced, above which friction sets in \cite{landau_theory_1941}. While in a BEC a superfluid phase was found \cite{raman_evidence_1999}, the transition is not as sharp as in the classical argument, 
as previous works also discussed e.g. \cite{ianeselli_beyond_2006,baym_landau_2012,singh_probing_2016}. The classical argument 
cannot be translated to a quantum system with a finite size wave function which has a 
distribution of velocities.

We can only expect frictionless dynamics for a velocity distribution 
smaller than the critical velocity $\Delta v < c_{\mathrm{crit}}=\sqrt{\frac{\mu}{m}}$.
Using Heisenberg's uncertainty relation, $\Delta v = \Delta p/M \ge \hbar/(2M\Delta x)$,
we see that the spread of the wave function is bounded by the
relation $\frac{\hbar}{2M\Delta x} \lesssim \sqrt{\frac{\mu}{m}}$. The size of
a wave packet in a harmonic oscillator of the order of $\Delta x \sim \sqrt{\frac{\hbar}{M\Omega}}$
such that the Landau criterion translates into 
%$of the particle has to have a spread small oNevertheless having fixed the the trapping frequency 
%$\omega$ and a damping dependency $J\left( \omega\right) $, we connect $v_{\mathrm{crit}}$ with a $\omega_{\mathrm{crit}}$, 
%using Heisenberg's uncertainty for a particle in a harmonic oscillator. We estimate $v_{\mathrm{crit}}$ using the Thomas-Fermi limit by $v_{\mathrm{crit}}= \frac{\mu}{m}$ and assume that particle velocities with $ \left|v\right| \simeq \frac{\sigma_p}{M}$ exceed the critical speed. By doing so, we find the following relation:
\begin{equation}
 \hbar\Omega \lesssim 2 \frac{M}{m}\mu\label{QLC},
\end{equation}
the condition for strongly suppressed damping.
While this argument is only qualitative, the $\Omega^4$ dependence of the damping rate for
precisely these frequencies $\Omega<\mu$ supports the quantum Landau criterion $(\ref{QLC})$.
Moreover, it provides an explanation for the missing sharp transition in contrast to the classical criterion, due Heisenberg uncertainty. A similar conclusion was also found for a constant 
motion in a homogeneous gas \cite{suzuki_creation_2014}. \\
\newline
%\hspace{5cm}
%\section{coherent heating}
$Coherent~heating$- 
After understanding the initial cooling dynamics, we now focus on the returning energy. Intuitively the process corresponds to a reheating of the ion, but in the following section we want to show that this process is more complex. \\
First of all it is important to note that not just the concept of a damping rate breaks down, it turns out it is more appropriate to stay in the time domain without referring to the spectral density $J\left( \Omega\right) $. We look at a simplified model for the correlation function $\alpha\left( t-s\right) $ for $ 0\le t \le 2T_{\mathrm{ret}}$:
\begin{align}
 \alpha \left( t-s\right) = 2 \gamma_1 \delta\left( t-s\right)  + \gamma_2 \delta\left( T_{\mathrm{ret}}-\left( t-s\right) \right) ,
\end{align}
mimicking the true $\alpha\left( t-s\right) $ from Figure (\ref{beta_pic}).
The initial decay is described by $\gamma_1$ and the strength of the return by $\gamma_2$, two fitting parameters. The solution to this equation can be found by changing into interaction picture and approximating the system for large frequencies \cite{waurick_$g$-convergence_2014} and generic initial conditions. In this manner the oscillations are averaged out, but still simulating the exponential energy loss. The solution to the problem can then be found as:
\small
\begin{align}\label{delay_eq}
\tilde Q\left(t\right) = &Q_0\ee^{-\gamma_1 t}~~~~~~~~~~~~~~~~~~~~~~~~~~~~~~~~~~~~~~~~~~~~~~~~~~~0\le t\le T_{\mathrm{ret}} \\\nonumber
\tilde Q\left(t\right) = &Q_0\ee^{-\gamma_1 t}- \gamma_2\left\{Q_0\cos\left(\omega T_{\mathrm{ret}}\right)-\frac{V_0}{\omega}\sin\left(\omega T_{\mathrm{ret}}\right)\right\}\\ \nonumber
&\ee^{-\gamma_1\left(t-T_{\mathrm{ret}}\right)}\left(t-T_{\mathrm{ret}}\right)~~~~~~~~~~~~~~~~~~~~~~~~~~~~~T_{\mathrm{ret}} \le t \le 2T_{\mathrm{ret}}
\end{align}
\normalsize
$\tilde Q\left(t\right) $ is illustrated in Figure $\ref{Klassik_50}$, where $\gamma_1$ is chosen as $\gamma\left(\Omega \right)$ and $\gamma_2$ was fitted. The solution shows the principal behavior of the ion including the coherent heating. The heating or cooling effect can be explained by the prefactor $\left\{Q_0\cos\left(\omega T_{\mathrm{ret}}\right)-\frac{V_0}{\omega}\sin\left(\omega T_{\mathrm{ret}}\right)\right\}$ appearing in solution (\ref{delay_eq}). If it is negative it results in heating and vice versa. Coherent heating can be observed for the whole spectrum, albeit the effect is too small to be significant for the low and very high frequency regime. Remarkably a small change in parameters can lead to a more dramatic effect. Also the damping strength can increase by orders of magnitude rapidly, making the numerical calculations 
more challenging. Besides, by tuning the physical parameters one could achieve striking cooling rates with high frequency changes.
\newline
%\section{Conclusion}

$Conclusion$- 
In this article we studied the joined dynamics of a BEC with a single ion. For describing the dynamics we used the bath correlation function and spectral density for solving the equations of motion.We found that the exact correlation function is most easily determined in the time domain.\\
Initially the ion is cooled by the BEC with a damping rate, depending on its trapping frequency $\omega$. The bare trapping frequency is increased by the interaction with the gas.
We identified a superfluid regime for small frequencies $\hbar \Omega \lesssim \mu$ with a superohmic suppression $\sim \Omega^4$ of the damping rate. For high $\omega$ the interaction potential determines the decay of the cooling rate. After $T_{\mathrm{ret}}$ the energy returns leading to coherent heating, which, counterintuitively, may result in additionally cooling depending on the fine tuning of the frequency.\\
%\section{Acknowledgments}
We thank the DFG (Grand No. STR $418/4-1$) and the IMPRS Dresden for financial support.

 %

%\bibliography{paper3.bib}
% \bibliographystyle{unsrtnat} 
% \bibliographystyle{extra.bst}
%\bibliographystyle{unsrt}

\bibliographystyle{apsrev4-1}

\end{document}